\documentclass[12pt,preprint]{aastex}

\usepackage{psfig}
 
\begin{document} 
\title{Polarization observations of nine southern millisecond pulsars}
\author{R. N. Manchester}
\affil{Australia Telescope National Facility, CSIRO, PO Box 76,
        Epping NSW 1710, Australia}
\email{Dick.Manchester@csiro.au}
\and
\author{J. L. Han} 
\affil{National Astronomical Observatories, Chinese Academy of 
	Sciences, Beijing 100012, China}
\email{hjl@bao.ac.cn}

\begin{abstract} 
Mean pulse profiles and polarization properties are presented for nine
southern pulsars. The observations were made using the Parkes radio
telescope at frequencies near 1330 MHz; three of the nine pulsars were
also observed at 660 MHz. A very high degree of circular polarization
was observed in PSR J1603$-$7202. Complex position angle variations
which are not well described by the rotating-vector model were observed in PSRs
J2124$-$3358 and J2145$-$0750, both of which have very wide
profiles. Rotation measures were obtained for all nine pulsars, with
two implying relatively strong interstellar magnetic fields. 
\end{abstract}

\keywords{Pulsars: general --- polarization --- ISM: magnetic field}

\section{Introduction}
Observations of the polarization of mean pulse profiles have given
much information about the pulsar radio emission mechanism. Pulsars
often have very high linear polarization and the sweep of position
angle across the pulse is generally well described by the
rotating-vector model \citep{rc69a,kom70}. This and the bilateral
symmetry of many pulse profiles led to a model in which the emission
is beamed into a hollow cone centered on the magnetic axis. With
observations of an increasing number of pulsars it soon became clear
that the mean pulse profiles of many pulsars were more complex. The
frequent occurence of a central emission peak \citep{bac76} led to the
idea of a `core' component in contrast to the outer `conal'
components. Further observations showed that many pulsars had multiple
conal components leading to ideas of multiple cones
\citep{ran93,kwj+94,md99} or patchy beams \citep{lm88,hm01}. Core
emission tends to have a steeper spectrum than conal emission and a
higher degree of circular polarization, often with a reversal of sense
near the profile center \citep{ran83,lm88}, although reversing
circular polarization is sometimes observed in conal emission
\citep{hmxq98}. The apparent width of both core and conal beams is a
function of pulse period ($P$), with an approximate $P^{-1/2}$
dependence \citep{ran90,big90b}.

With the discovery of millisecond pulsars (MSPs), the parameter ranges
available for study of pulse emission processes was greatly extended.
MSPs have much weaker surface magnetic fields than `normal' pulsars,
with $B_0$ typically a few by $10^8$~G, compared to $10^{12}$~G for
normal pulsars. They also have much smaller magnetospheres,
conventionally defined to be limited by the light cylinder at radius
$R_{LC} = cP/2\pi$, where $P$ is the pulsar period and $c$ is the
velocity of light. For the fastest MSP, PSR B1937+21, $R_{LC}$ is only
74 km, or about ten times the neutron star radius. One would therefore
expect the radio emission from MSPs to be rather different to that
from normal pulsars. MSP pulse profiles are generally wider (in pulse
phase) than those of normal pulsars, which is not surprising in view
of the wider angle subtended by the `open' field lines which penetrate
the light cylinder and define the extent of the polar cap. Also, MSP
profiles have different frequency evolution characteristics compared
to most normal pulsars. In particular, the component separations are
largely independent of frequency \citep{kll+99}. It is not clear
that the distinctions and relationships between `core' and `conal'
emission seen in longer-period pulsars apply to MSPs. For example, the
strong component in PSR J0437$-$4715 is clearly central to the
emission beam but has a flatter spectrum than the outer components
\citep{nms+97}.

Despite these differences, there are many similarities in the radio
emission properties of MSPs and normal pulsars. If the greater width
of MSP profiles is ignored, the shapes of MSP mean pulse profiles are
very similar to those of normal pulsars. However, among the MSPs,
there is a greater proportion of `interpulse' profiles, with two main
regions of pulse emission separated by approximately $180\degr$ of
pulse phase or longitude.  MSP profiles often appear to have more
components than normal pulsars, but Kramer et
al. (1998)\nocite{kxl+98} have argued that this is simply because MSP
components are more obvious because of their wider spacing. On
average, the radio emission of MSPs has a spectral index which is
only marginally steeper than that of normal pulsars and may in fact be
the same \citep{tbms98,kxl+98}. MSPs are typically somewhat less
luminous than normal pulsars but there is a large overlap in the
luminosity distributions. Finally, the polarization properties of MSPs
and normal pulsars are remarkably similar, with both often having high
fractional linear polarization and generally smaller levels of
circular polarization. Orthogonal jumps in position angle are seen in
both classes of pulsar (e.g. Stinebring et al. 1984\nocite{scr+84};
Thorsett \& Stinebring 1990\nocite{ts90}).  Variations of position
angle (PA) through MSP profiles are often more complex than is typical
for normal pulsars. Despite this, in many cases the PA can be
approximately fitted by the simple rotating vector model (RVM) which
applies to most normal pulsars, at least once orthogonal jumps are
taken into account (e.g. Arzoumanian et
al. 1996)\nocite{aptw96}. These similarities leave little doubt that
the basic radio emission mechanism is the same in normal pulsars and
MSPs.

Extensive studies of the polarization properties of MSPs have
previously been made by Thorsett \& Stinebring (1990)\nocite{ts90}:
PSRs B1937+21, B1953+29 and B1957+20 at several frequencies between
430 and 2830 MHz; Navarro et al. (1997)\nocite{nms+97}: PSR
J0437$-$4715 at 438, 660 and 1512 MHz; Arzoumanian et
al. (1996)\nocite{aptw96}: PSR B1534+12 at 430 MHz; Xilouris et
al. (1998)\nocite{xkj+98}: 24 pulsars at 1410 MHz; Stairs, Thorsett \&
Camilo (1999)\nocite{stc99}: 19 pulsars at 410, 610 and/or 1414 MHz;
and Lommen et al. (2000)\nocite{lzb+00}: PSR J0030+0451 at 430
MHz. All observations except those of PSR J0437$-$4715 by Navarro et
al. were made using northern hemisphere telescopes, so there is little
information on the polarization of the southern MSPs. Also, there are
significant discrepancies in the results of different groups for some
pulsars. For example, results from Xilouris et al. (1998) and Stairs
et al. (1999) for PSRs J1022+1001, J1713+0747 and J2145-0750 at
frequencies near 1400 MHz differ substantially.\footnote{Stairs et
al. (1999) use a convention for sense of position angle which is
opposite to the astronomical convention used by other authors.} It is
important to understand whether these differences are due to time
variations such as mode changing \citep{kxc+99} or problems with data
processing and/or calibration. Furthermore, in some previous studies
(e.g. Stairs et al. 1999) the position angle scale is not
absolute. Absolute position angles are important for Faraday rotation
studies and in comparisons with other properties. Important examples
are the comparison of the pulsar rotation axis direction implied by
radio polarization measurements with that deduced from X-ray
observations and with the direction of the pulsar proper
motion. \citep{hgh01,rd01}.

With their high linear polarization, it is relatively
easy to determine rotation measures (RMs) for most pulsars. Combined with
the dispersion measure (DM), pulsars give a direct measure of the mean
line-of-sight magnetic field between the Sun and the pulsar, weighted
by the local electron density. Since the DM also provides an estimate
of the pulsar distance, pulsars are a valuable probe of the Galactic
magnetic field (e.g. Han et al. 2002)\nocite{hmlq02}. Most MSPs are
relatively close to the Sun, so they provide information on
conditions in the local region of the Galaxy.

We have used the Parkes 64-m telescope of the Australia Telescope
National Facility and the Caltech pulsar correlator to make
observations of nine southern MSPs at 20~cm and, for three of these,
at 50~cm. Five of these pulsars have no previously published
polarization observations and none have previously published rotation
measures. Details of the observing system and observational parameters
are given in \S~2. The polarization and RM results are presented in
\S~3 and \S~4 respectively, and in \S~5 we discuss some implications
of our polarization measurements.

\section{Observations}
Observations were made in three sessions, 1996 August 31 -- September
5, 1999 December 12 -- 17, and 2000 December 14 -- 19 using the Parkes
64-m radio telescope. The 20-cm observations were made using the
central beam of the Parkes multibeam receiver \citep{swb+96} in the
1999 and 2000 sessions with a central frequency of 1318.5 MHz. This
system has cryogenically cooled preamplifiers, giving a system
temperature of about 21~K or about 29 Jy, and orthogonal linear feeds
with a calibration signal injected at angle of $45\degr$ to the
feed probes. Because this system has significant cross-coupling
between the feed probes, all observations were made at two feed angles
at $\pm 45$\degr. These measurements were summed after compensation
for the feed rotation, thereby eliminating most of the effects of feed
cross-coupling. The 50-cm observations had a central frequency of
660 MHz and were were made in the 1996 and 2000 sessions
using a receiver with cryogenically cooled preamplifiers and a
cavity-backed disk feed, again with a calibration signal injected at
45\degr. Observations with this system were made at fixed feed angle;
in this case the feed cross-coupling is small and did not
significantly affect the polarization results. The flux scale was
established using observations of Hydra A, assumed to have flux
densities of 85 and 45 Jy at 660 and 1320 MHz, respectively.

The Caltech correlator system \citep{nav94} was used to form the
polarization products and to bin the signals synchronously with the
pulsar period. For both 660 and 1320 MHz, the signals were
up-converted to 1580 MHz, band-limited to 32 or 128 MHz respectively
and power levels adjusted for 2-bit digitization at 256 MHz. The
correlator gives 128 lags in each of four polarization channels and
folds the data synchronously with the pulsar period with 1024 bins per
period. After on-line folding, typically for 90 seconds, the 512
sub-integration profiles were transferred to a workstation for further
processing. Data in each phase bin were transformed to the frequency
domain and narrow-band interference rejected. Using a pulsed
calibration observation made immediately before each pulsar
observation, the data were corrected for differential delays between
the two polarization channels and calibrated into mJy units. After
conversion to Stokes parameters, the data were corrected for
parallactic angle rotation and ionospheric Faraday rotation and
dedispersed to form 16 to 64 frequency sub-bands, depending on the
pulsar DM.

In off-line analysis, the orthogonal observation pairs for 1320 MHz
were summed and the data were summed in time to form Stokes-parameter
profiles for each sub-band. In most cases several
observation pairs were summed to form the final
profiles. Because of low system gain, some sub-bands at one or both
ends of the bandpass were normally rejected, changing the effective
center frequency. We then summed the data in frequency for a range of
trial RMs, (normally in a range of $\pm$1500 rad~m$^2$ with a step of
25 rad~m$^2$) to search for a peak in the linearly polarized intensity
$L = (Q^2 + U^2)^{1/2}$. Stokes parameter profiles were then formed
for the upper and lower halves of the bandpass, correcting for the
approximate RM determined in the previous stage of analysis. The best
estimate of pulsar RM was then obtained by taking weighted means of
position-angle (PA) differences for bins where the uncertainty in the
PA difference was less than $10\degr$. Finally, a single set of
Stokes-parameter profiles was formed for each observation.

Table~\ref{tb:obspar} lists the basic observational parameters for the
nine observed MSPs. Columns are, in order: pulsar name based on
J2000 coordinates, pulse period, DM, year(s) in which the observations
were made, effective center frequency and bandwidth after sub-band
rejection, number of bins across the pulse period and the total
observation time. 
\begin{table*}
\begin{minipage}{135mm}
\begin{small}
\caption{Observational parameters for southern millisecond pulsars}\label{tb:obspar}
\begin{tabular}{lrccccccl}
\hline
\multicolumn{1}{c}{PSR}&
\multicolumn{1}{c}{$P$} &
\multicolumn{1}{c}{DM} &
\multicolumn{1}{c}{Obs.} &
\multicolumn{1}{c}{Ctr Freq.} &
\multicolumn{1}{c}{BW} &
\multicolumn{1}{c}{Nr of} &
\multicolumn{1}{c}{Obs. time} \\
 &
\multicolumn{1}{c}{(ms)}&
\multicolumn{1}{c}{(cm$^{-3}$pc)}&
\multicolumn{1}{c}{Date} &
\multicolumn{1}{c}{(MHz)} &
\multicolumn{1}{c}{(MHz)} &
\multicolumn{1}{c}{Bins} &
\multicolumn{1}{c}{(s)} \\
\hline
J0613$-$0200 & 3.062 & 38.79 & 99+00 & 1335 & 96 & 64 & 3960 \\
J0711$-$6830 & 5.491 & 18.41 & 99+00 & 1335 & 96 &128 & 5580 \\
J1045$-$4509 & 7.474 & 58.15 & 99    & 1335 & 96 &128 & 2700 \\
J1603$-$7202 &14.842 & 38.05 & 96    &  661 & 24 &128 & 1800 \\
            &       &       & 99    & 1327 & 96 &256 & 2700 \\
J1623$-$2631 &11.076 & 62.86 & 00    & 1331 & 96 &128 & 1080 \\
J1643$-$1224 & 4.622 & 62.41 & 00    & 1331 & 96 & 64 & 1080 \\
J2124$-$3358 & 4.931 &  4.62 & 00    &  659 & 24 &256 & 2160 \\
            &       &       & 99+00 & 1327 & 80 &128 &11520 \\
J2129$-$5721 & 3.726 & 31.85 & 96    &  659 & 24 & 64 & 2880 \\
            &       &       & 99    & 1331 & 96 &128 & 2880 \\
J2145$-$0750 &16.052 &  9.00 & 00    & 1335 & 96 &256 & 2160 \\
\hline
\end{tabular}\\
\end{small}
\end{minipage}
\end{table*}
 
\section{Results}

The results are presented in terms of Stokes $I$ (total intensity),
linearly polarized intensity $L$ and Stokes $V$, defined in the sense
$I_L - I_R$, where $I_L$ and $I_R$ are the left- and right-circularly
polarized intensities respectively according to the IRE
definition. The linear position angle is measured from north toward
east (counter-clockwise) on the sky, in the normal astronomical
convention. Table~\ref{tb:polpar} gives a summary of the
results. Columns are, in order: pulsar name, mean flux density,
fractional linear polarization, fractional net circular polarization,
fractional absolute circular polarization, estimated error in the
fractional polarizations, pulse width at 50\% of the peak in degrees
(where the pulse period = 360\degr) and milliseconds, and pulse width
at 10\% of the peak in degrees and milliseconds. The mean flux density
$S$ is the sum of $I$ across the profile divided by the number of bins
per period, and the fractional polarizations are the sum of the
polarization quantity ($L$, $V$ or $|V|$) across the pulse profile
divided by the sum of $I$ across the profile.

\begin{table*}
\begin{minipage}{165mm}
\begin{small}
\caption{Polarization and pulse width parameters for southern millisecond pulsars }\label{tb:polpar}
\begin{tabular}{lcrccccrrcrrr}
\hline
\multicolumn{1}{c}{PSR}&
\multicolumn{1}{c}{Freq.} &
\multicolumn{1}{c}{$S$}&
\multicolumn{1}{c}{$\langle L\rangle/S$}&
\multicolumn{1}{c}{$\langle V\rangle /S$}&
\multicolumn{1}{c}{$\langle|V|\rangle/S$}&
\multicolumn{1}{c}{$\sigma$ } &
\multicolumn{2}{c}{ W$_{50}$ } & &
\multicolumn{2}{c}{ W$_{10}$ } 
\\ \cline{8-9} \cline{11-12}
 &
\multicolumn{1}{c}{(MHz)}&
\multicolumn{1}{c}{(mJy)}&
\multicolumn{1}{c}{(\%)} &
\multicolumn{1}{c}{(\%)} &
\multicolumn{1}{c}{(\%)} &
\multicolumn{1}{c}{(\%)} &
\multicolumn{1}{c}{($\degr$)} &
\multicolumn{1}{c}{(ms)} & &
\multicolumn{1}{c}{($\degr$)} &
\multicolumn{1}{c}{(ms)} 
\\
\hline
J0613$-$0200 & 1335 & 4.6 & 15 &   2 &  4 & 1 &  67 & 0.57&&130 & 1.11 \\
J0711$-$6830 & 1335 & 2.1 & 13 &$-$16& 19 & 2 & 125 & 1.90&&175 & 2.68 \\
J1045$-$4509 & 1335 & 2.4 & 14 &  14 & 17 & 2 &  40 & 0.82&& 78 & 1.62 \\
J1603$-$7202 &  661 &21.6 & 13 &  30 & 30 & 1 &  38 & 1.57&& 63 & 2.59 \\
            & 1327 & 7.0 & 13 &  29 & 30 & 1 &  33 & 1.34&& 49 & 2.02 \\
J1623$-$2631 & 1331 & 1.4 & 34 &   1 & 18 & 3 &  19 & 0.57&& 88 & 2.71 \\
J1643$-$1224 & 1331 & 5.1 &  7 & $-1$& 11 & 1 &  47 & 0.60&&114 & 1.47 \\
J2124$-$3358 &  659 & 130 & 41 & $-5$&  9 & 1 & 169 & 2.32&&330 & 4.52 \\
            & 1327 & 7.9 & 29 & $-1$&  6 & 1 &  37 & 0.51&&278 & 3.82 \\
J2129$-$5721 &  659 & 6.8 & 50 &$-28$& 30 & 2 &  46 & 0.47&&113 & 1.17 \\
            & 1331 & 1.4 & 45 &$-28$& 31 & 2 &  30 & 0.31&& 83 & 0.86 \\
J2145$-$0750 & 1335 &16.6 & 15 &   7 &  9 & 1 & 8.5 & 0.38&& 94 & 4.20 \\
\hline
\end{tabular}
\end{small}
\end{minipage}
\end{table*}

In the following subsections we present polarization profiles and discuss
each pulsar in turn. 

\subsection{PSR J0613$-$0200}
\begin{figure}
\psfig{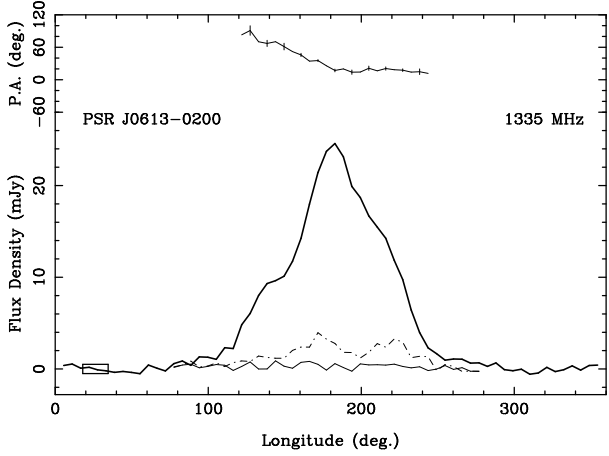}
\caption{Polarization profiles for PSR J0613$-$0200 at 1335 MHz. In
this and subsequent figures of this type the lower part gives the pulse
profiles for total intensity $I$ (thick line), linearly polarized
intensity $L$ (dot-dashed line) and circularly polarized intensity $V$
(thin line). The abscissa is degrees of longitude where
360\degr~equals the pulse period.  The error box on the left-hand side
of the profile has a width equal to the profile resolution
(including dispersion smearing) and an amplitude of twice the baseline
rms noise (i.e. $\pm 2\sigma$).  In the upper part, the position angle of the
linearly polarized emission is plotted where its uncertainty is less than
$10\degr$. Error bars ($\pm 2\sigma$) are given on every second
point.}
\label{fg:0613}
\end{figure}
Figure~\ref{fg:0613} shows that this pulsar has a wide profile with
three main components. Because of the short pulsar period and
relatively high DM (Table~\ref{tb:obspar}) our observations are
significantly smeared so the three components are not clearly
resolved. However, the profile is very similar to that observed by
\citet{xkj+98} at a similar frequency. \citet{stc99} observe a quite
different profile at 410 and 610 MHz, with the trailing component much
stronger than the other two, showing that this component has a
steeper spectrum. This frequency evolution is also consistent with the
pulse profiles presented by \citet{bbm+97}. At 610 MHz, \citet{stc99}
observe a reversal of circular polarization under this trailing
component. This and the steep spectrum suggest that it may be `core'
emission despite its trailing location.

We observe weak linear polarization with a slow decrease in position
angle across the profile and no significant circular
polarization. Given the different frequency and time resolution,
these results are not in disagreement with those observed by
\citet{stc99}. \citet{xkj+98} did not show linear polarization in the
figure for this pulsar, but give a fractional linear polarization of
26\% and fractional circular polarization of $-12$\% in their Table 1,
both much higher than we observe (Table~\ref{tb:polpar}).

\subsection{PSR J0711$-$6830}
\begin{figure}
\psfig{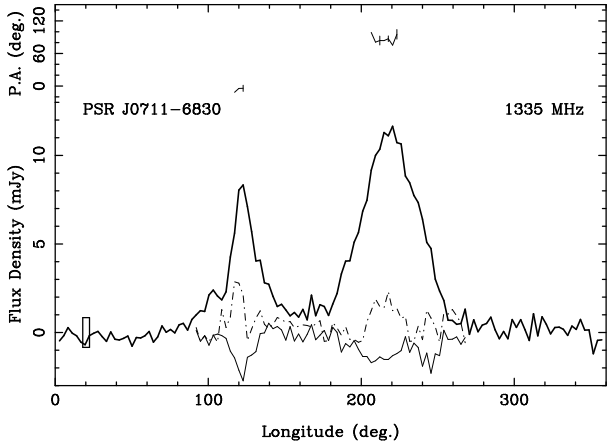}
\caption{Polarization profiles for PSR J0711$-$6830 at 1335 MHz. See
  Figure~\ref{fg:0613} for details.}
\label{fg:0711}
\end{figure}
The pulse profile for this pulsar (Figure~\ref{fg:0711}) has two broad
and widely separated components connected by a bridge of
emission. Profiles at frequencies between 430 and 1500 MHz published
in the discovery paper \citep{bjb+97} show the leading component has a
steeper spectrum than the trailing component. Our results show
significant right-circular polarization across both components and
linear polarization near the center of each component. Position 
angles differ by about $90\degr$ between the two components, but it is
not clear if there is continuity between them. There are no previously
published polarization measurements.

\subsection{PSR J1045$-$4509}
\begin{figure}
\psfig{file=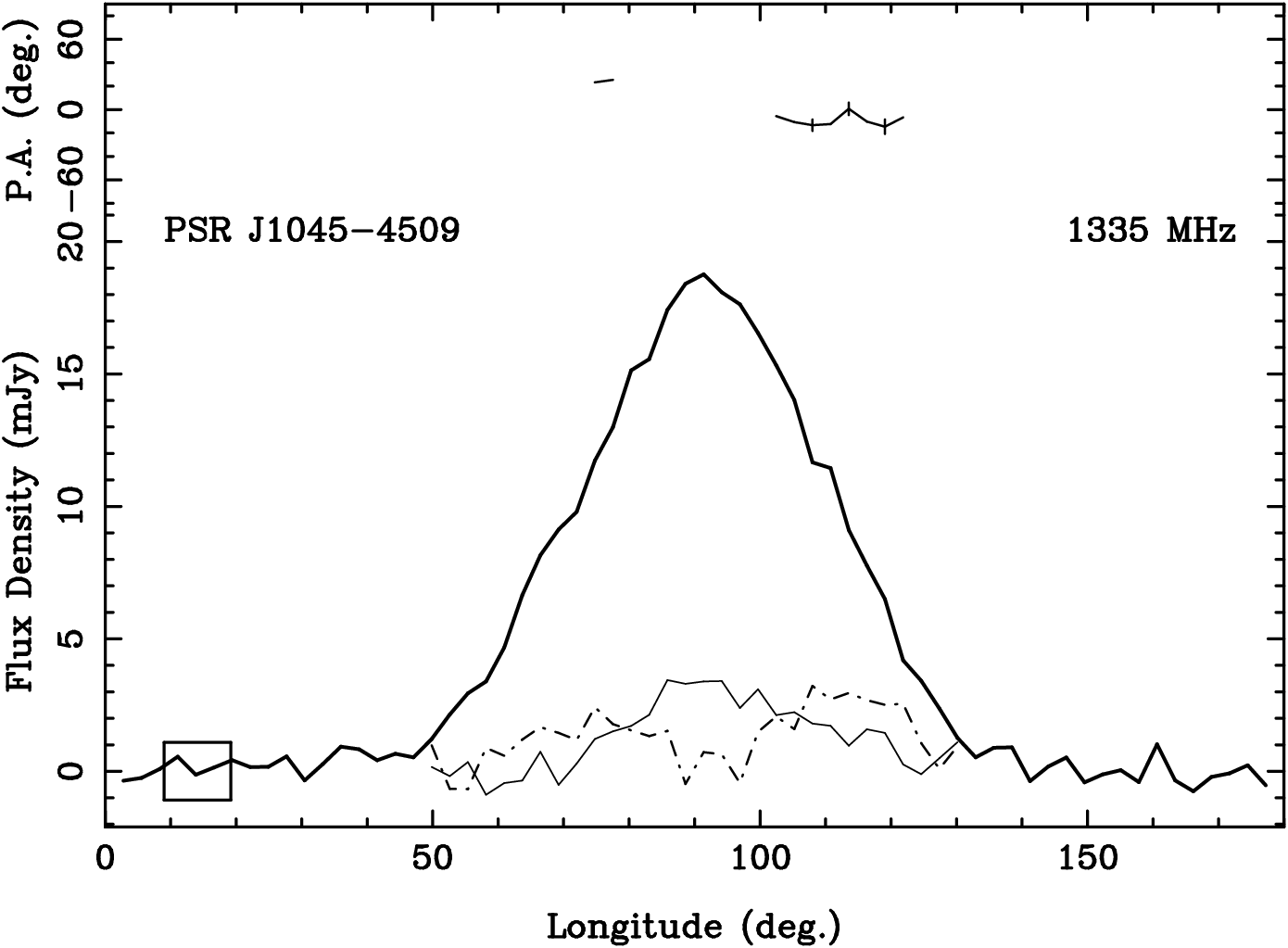,width=80mm,angle=270}
\caption{Polarization profiles for PSR J1045$-$4509 at 1335 MHz. See
  Figure~\ref{fg:0613} for details.}
\label{fg:1045}
\end{figure}
Figure~\ref{fg:1045} shows that this pulsar has a triangular shaped
profile with a 50\% width of 40\degr~(Table~\ref{tb:polpar}). Most of
the profile has weak left-circular polarization. The trailing part
has significant linear polarization with approximately constant
position angle. Again, there are no previously published polarization
measurements.

\subsection{PSR J1603$-$7202}
\begin{figure*}
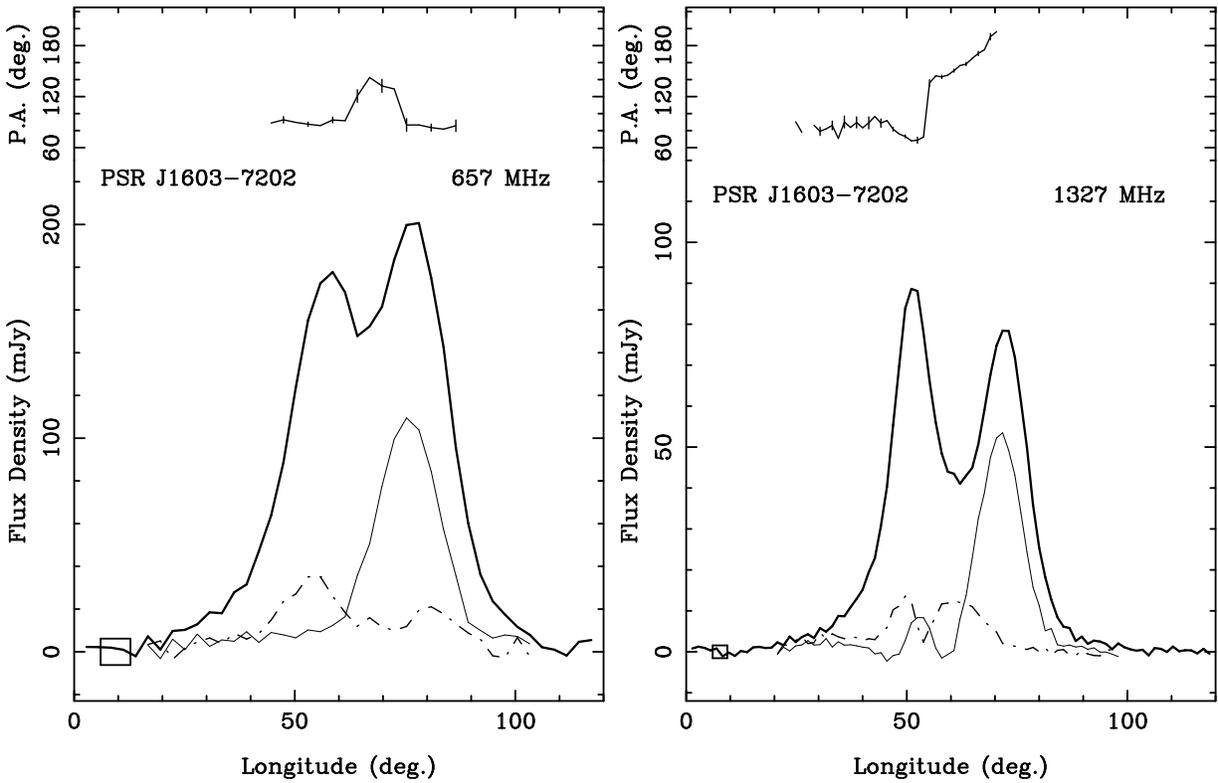

\begin{tabular}{ll}
\mbox{\psfig{file=f4a.eps,width=80mm,angle=0}} 
\mbox{\psfig{file=f4b.eps,width=80mm,angle=0}} 
\end{tabular}
\caption{Polarization profiles for PSR J1603$-$7202 at 657 and 1327
MHz. See Figure~\ref{fg:0613} for details. }
\label{fg:1603}
\end{figure*}

The only previously published profile for this pulsar is from the
discovery paper \citep{llb+96} at 436 MHz which shows two barely
resolved components. Figure~\ref{fg:1603} shows that the components
become more separated at higher frequencies (in contrast to the
frequency dependence observed in most normal pulsars) and that the
leading component has a flatter spectrum.

The polarization properties of this pulsar, specifically the trailing
component, are extraordinary. This component is one of the most highly
circularly polarized broad-band radio continuum sources known, with a
fractional circular polarization of more than 50\% at 657 MHz and more
than 65\% at 1327 MHz. Weaker but still significant circular
polarization is seen in the leading component, apparently with two
sense reversals.

There is a clear but slightly smeared orthogonal flip in the linear
polarization on the trailing edge of the first component at 1327 MHz
with an accompanying dip in the linearly polarized intensity
indicating overlap of two orthogonally polarized modes. Even when this
is allowed for, the position angle variation across the pulse deviates
from the simple RVM form. At 657 MHz, there may be two orthogonal
flips of PA -- it is difficult to tell because of the dispersion
smearing. If so, the PA is approximately constant across the pulse.

\subsection{PSR J1623$-$2631 (PSR B1620$-$26)}
\begin{figure}
\psfig{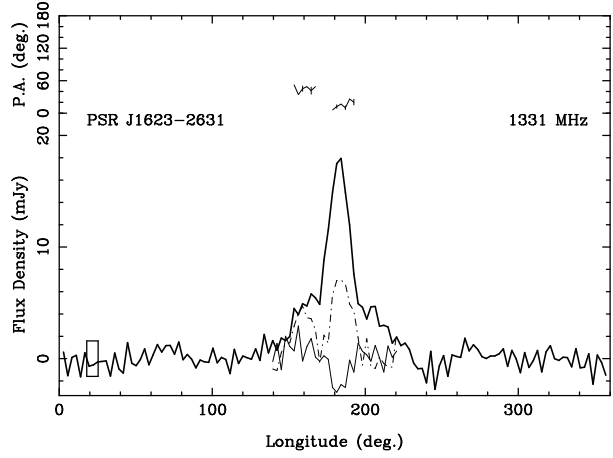}
\caption{Polarization profiles for PSR J1623$-$2631 at 1331
MHz. See Figure~\ref{fg:0613} for details. }
\label{fg:1623}
\end{figure}
This three-component pulsar has high linear polarization in the
leading component and moderate linear and right-circular polarization
in the central component (Figure~\ref{fg:1623}). There is some
indication of an increasing PA with an orthogonal jump between the two
components. These results are fully consistent with the results at
1408 MHz by \citet{gl98} and those at 610 MHz by \citet{stc99}, given
the opposite sense of PA definition. In contrast, \citet{xkj+98} see a
weakly polarized first component, decreasing PA through the central
component, no orthogonal jump and left-circular polarization in the
central component.

\subsection{PSR J1643$-$1224}
\begin{figure}
\psfig{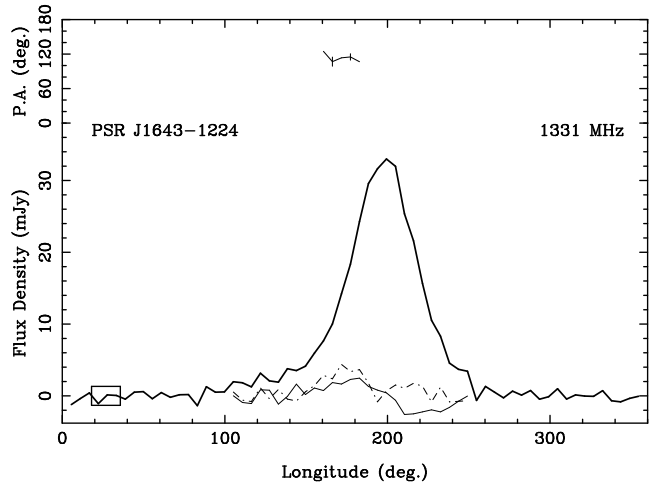} 
\caption{Polarization profiles for PSR J1643$-$1224 at 1331
MHz. See Figure~\ref{fg:0613} for details. }
\label{fg:1643}
\end{figure}
PSR J1643$-$1224 has a broad profile of almost gaussian shape except
for an extended ramp of emission at the leading edge
(Figure~\ref{fg:1643}). There is some linear polarization in the first
half of the pulse and weak circular polarization through the pulse
with a sign reversal from left to right near the center. Within our
uncertainties, the PA is constant across the linearly polarized part.

The 610-MHz data presented by \citet{stc99} are consistent with those
in Figure~\ref{fg:1643} except that they see no right-circular
emission in the trailing half of the pulse. At 1410 MHz \citet{xkj+98}
find linear polarization near the pulse peak with a slow decrease of
PA and right-circular polarization through most of the profile.

\subsection{PSR J2124$-$3358}
\begin{figure*}
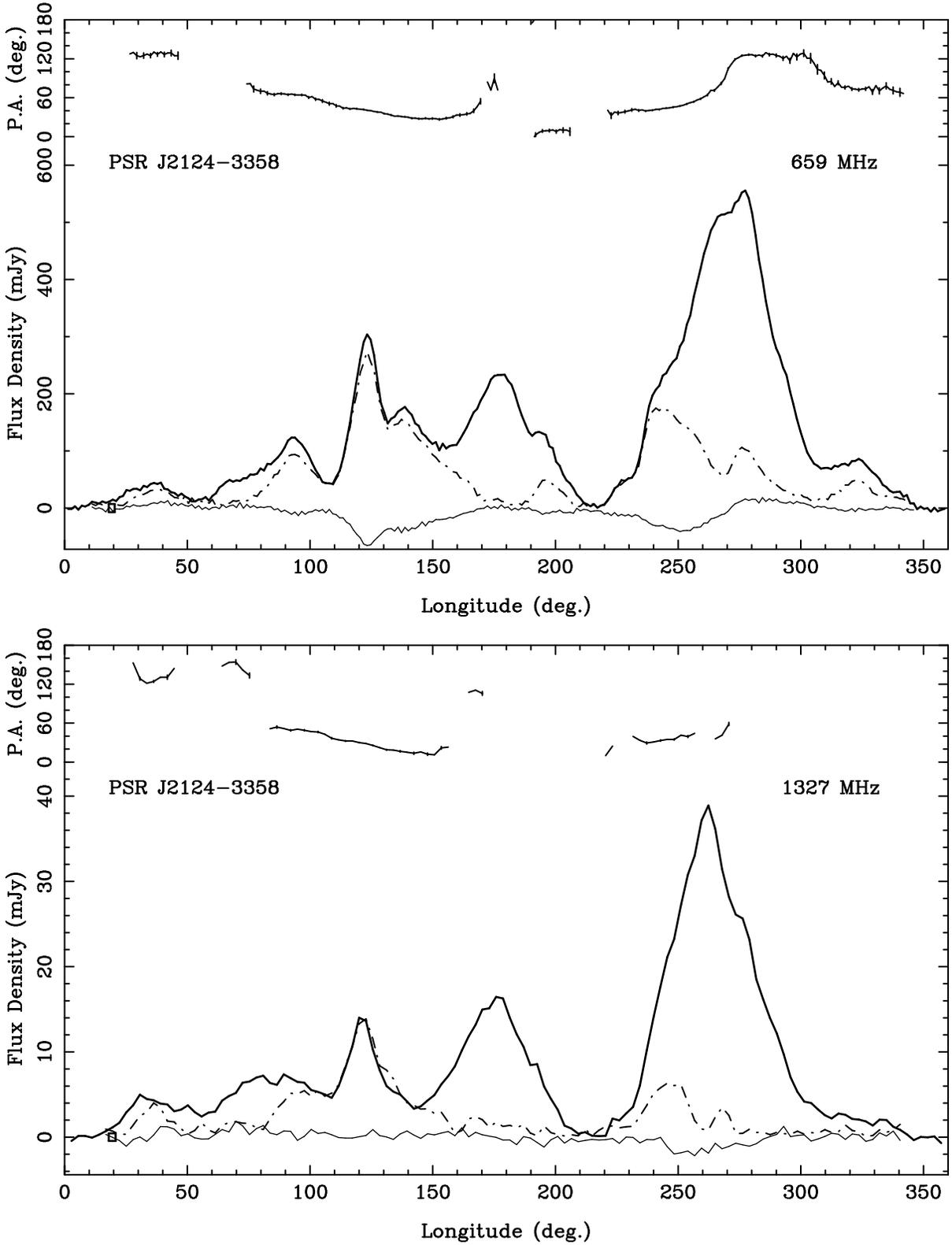

\begin{tabular}{c}
\mbox{\psfig{file=f7a.eps,width=160mm,angle=270}} \\
\mbox{\psfig{file=f7b.eps,width=160mm,angle=270}} 
\end{tabular}
\caption{Polarization profiles of PSR J2124$-$3358 at 659 MHz and 1327
MHz. See Figure~\ref{fg:0613} for details.}
\label{fg:2124}
\end{figure*}

Figure~\ref{fg:2124} presents the first polarization observations of
this isolated millisecond pulsar \citep{bjb+97}. The pulsar has an
incredibly complex pulse profile with emission over essentially the
whole pulse period. At least 12 pulse components can be identified in
the 659-MHz profile which has higher signal-to-noise ratio than the
1327 MHz one. In fact this pulsar very likely emits at all rotational
phases. This may be a serious problem for the polarization calibration
as the true baseline for the Stokes parameters cannot be
determined. We have taken the baseline to be the average level over
20\degr~of longitude at the minimum of the Stokes $I$ curve. If the
pulsar has significant emission at this minimum phase, the
polarization parameters derived here will be systematically biased,
especially where the fractional polarization is low
\citep[cf.][]{rr97}. The level of such emission can only be determined
by pulse-phase resolved interferometric measurements. It is possible
that this effect accounts for the small region near $130\degr$
longitude in the 1327-MHz profile where $L > I$, but this excess $L$
may simply be due to non-random noise. The fact that $L \leq I$
everywhere else gives us some confidence that the effect of unpulsed
emission, at both 659 and 1327 MHz, is small.

\begin{figure*}
\psfig{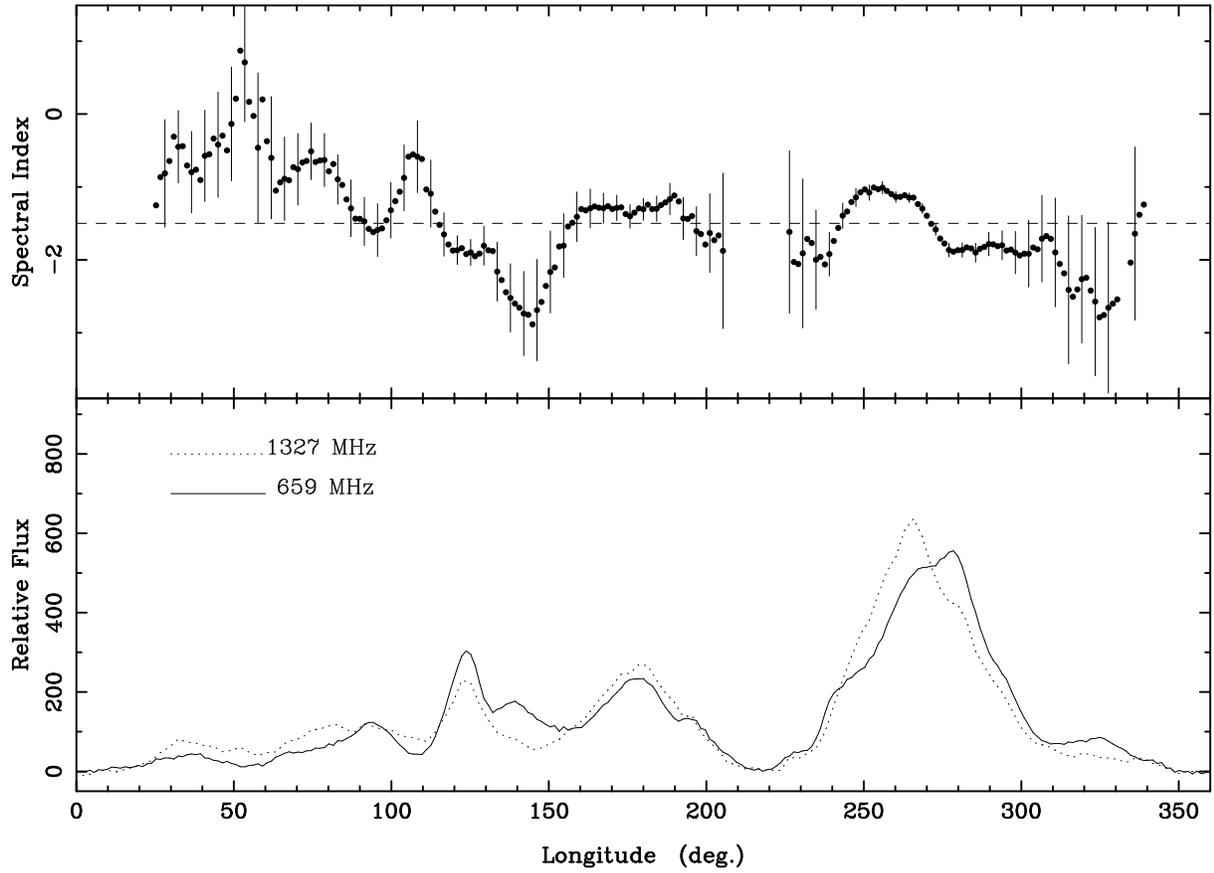}
\caption{Distribution of spectral index across the pulse profile for
PSR J2124-3358 derived from the 659 MHz and 1335 MHz data. }
\label{fg:2124_alpha}
\end{figure*}

The components vary in relative strength between 659 and 1327 MHz,
indicating a range of spectral indices. Figure~\ref{fg:2124_alpha}
shows the variations in spectral index across the pulse profile.
Absolute spectral indices cannot be estimated from these data as the
observed signal strength is strongly affected by interstellar
scintillation; in particular, the pulsar was exceptionally strong at
the time of the 659 MHz observations. The two Stokes $I$ profiles were
normalised to the same area and aligned in longitude by matching the
relatively isolated component near longitude $125\degr$ before
computing the differential spectral index. This value was then added
to the mean spectral index of $-1.5$ obtained by \citet{tbms98}. The
more positive values in regions of low-level emission may indicate a
larger amount of unpulsed emission at the lower frequency. However,
even discounting these regions, there is a remarkable variation in
spectral index across the profile. The weak component near 140\degr~
has a very steep spectrum, whereas the central features around
180\degr~ and the leading side of the strong peak have much flatter
spectra. This figure strongly suggests that individual components have
distinct spectral indices, with regions of varying spectral index
being due to overlapping components \citep[cf.][]{kwj+94}.

Portions of the profile are close to 100\% linearly polarized,
especially between 110\degr~and 135\degr~of longitude at both
frequencies and on the leading edge of the main component (220\degr~to
240\degr) at 659 MHz. On average, the fractional linear polarization
is much higher at the lower frequency (Table~\ref{tb:polpar}). The
variation of position angle across the profile is similar at the two
frequencies and extremely complex. There may be orthogonal transitions
at 1327 MHz around longitudes of $80\degr$ and $160\degr$. The PA 
variations at 659 MHz between longitudes of $170\degr$ and $190\degr$
should be viewed with caution as the fractional polarization is low
there. 

At both frequencies there appears to be weak right-circular
polarization in the leading part of the main peak, and at 659 MHz the
highly linearly polarized component around 120\degr~also has
right-circular polarization. Overall, the circular polarization is
much weaker than the linear, and weaker at 1327 MHz than at 659 MHz.

\subsection{PSR J2129$-$5721}
\begin{figure*}
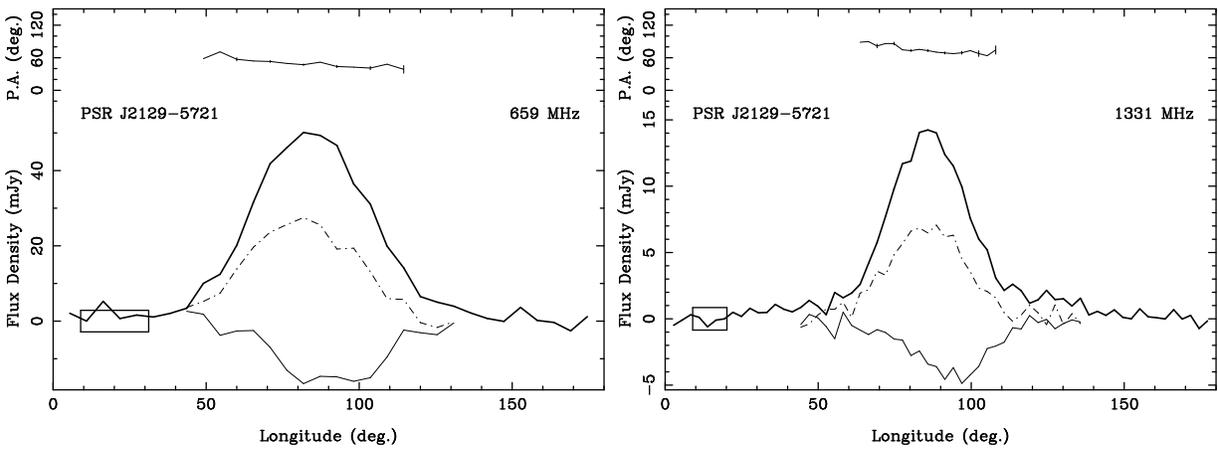

\begin{tabular}{ll}
\mbox{\psfig{file=f9a.eps,width=80mm,angle=270}} 
\mbox{\psfig{file=f9b.eps,width=80mm,angle=270}} 
\end{tabular}
\caption{Polarization profiles for PSR J2129$-$5721 at 659 and 1331
MHz. See Figure~\ref{fg:0613} for details. }
\label{fg:2129}
\end{figure*}
Figure~\ref{fg:2129} suggests that the mean pulse profile for this
pulsar consists of a single broad gaussian-shaped peak with extended
wings on both sides. However, especially at 659 MHz, the dispersion
smearing is large and so narrower features may exist in the
profile. Despite this smearing, it is clear that the profile has about
50\% linear polarization at both frequencies, with a very slow
decrease in PA across the profile. There is also relatively strong
right-circular polarization (about 30\%) through most of the profile, but with its
peak somewhat delayed with respect to the total
intensity profile.

\subsection{PSR J2145$-$0750}
\begin{figure*}
\psfig{file=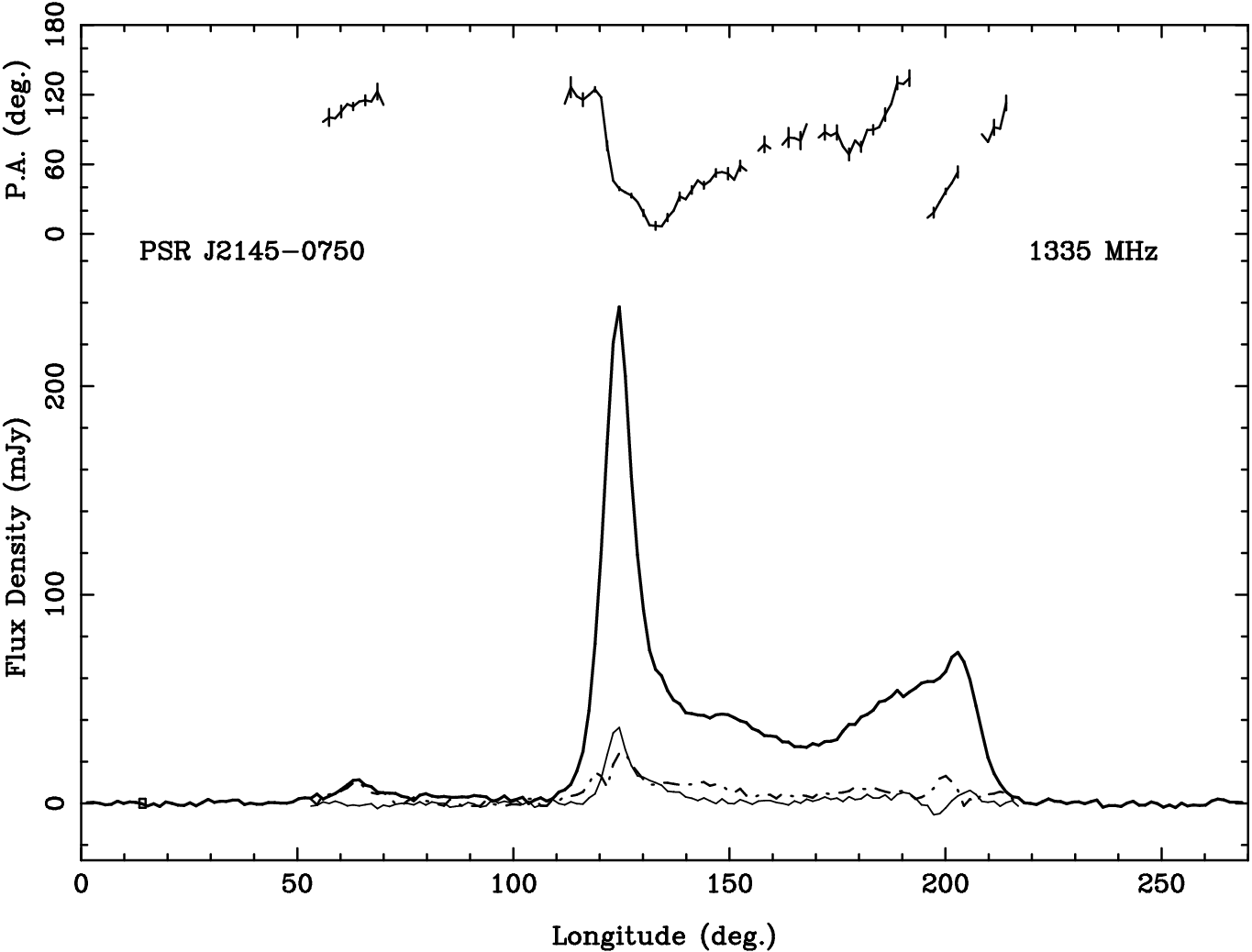,width=160mm,angle=270}
\caption{Polarization profiles for PSR J2145-0750 at 1335 MHz. See
Figure~\ref{fg:0613} for details. }
\label{fg:2145}
\end{figure*}
The profile for PSR J2145$-$0750 shown in Figure~\ref{fg:2145} was
obtained during a strong scintillation maximum and has a better
signal-to-noise ratio than previously published polarization profiles
\citep{xkj+98,stc99}. The total intensity profile at 1335 MHz consists
of three main components with the central one dominating at this
frequency. However, there is good evidence for at least two other
pulse components between the main peak and the trailing component.
Observations at frequencies between 102 MHz and 4850 MHz
\citep{kll+99} show that the trailing component has a much steeper
spectrum than the rest of the profile and dominates at low
frequencies. Figure~\ref{fg:2145} shows that there is a weak but
significant emission bridge between the leading and main component.

Although the weak leading component is essentially 100\% polarized,
linear polarization is relatively weak (but unambiguous) over most of
the profile; the mean fractional linear polarization
(Table~\ref{tb:polpar}) is 15\%. There appears to be a rapid swing of
position angle through the central component, but the higher
time-resolution observations of \citet{stc99} show that, at least at
lower frequencies, there is an orthogonal PA jump at this longitude.
There is also a rapid swing of PA in the trailing component, possibly
also with an orthogonal jump. Significant circular polarization is
observed in both components, with a sense reversal in the trailing
component. \citet{stc99} see a similar variation of circular
polarization at 410 and 610 MHz, except that the sense of the sign
reversal in the trailing component (left to right) is opposite to that
seen in Figure~\ref{fg:2145}. This may indicate a frequency evolution
of the circular polarization behaviour at this longitude. Their
1414-MHz polarization profiles are consistent with those shown in
Figure~\ref{fg:2145}. \citet{xkj+98} also see strong linear
polarization with an increasing PA in the weak leading component at
1410 MHz. However, their results for the stronger central and trailing
half of the profile, with relatively strong linear polarization
mimicing the total intensity profile and approximately constant PA,
are inconsistent with those in Figure~\ref{fg:2145}.  \citet{xkj+98}
also found a peak in circularly polarized intensity in the central
component, but it is of opposite sign to that in Figure~\ref{fg:2145}.

\section{Rotation measures}
Derived rotation measures for the nine MSPs are given in
Table~\ref{tb:rm} along with the pulsar Galactic longitude, Galactic
latitude and distance estimated from the DM using the NE2001 
Galactic electron density model \citep{cl02}. The mean line-of-sight component of
the interstellar magnetic field, weighted by the local electron
density, given by
\begin{equation}
\langle B_{||}\rangle = 1.232\,\frac{\rm RM}{\rm DM}\; \mu{\rm G},
\end{equation}
where RM is in units of rad m$^{-2}$ and DM is in cm$^{-3}$
pc, is listed in the final column of the table. A positive $\langle
B_{||}\rangle$ corresponds to a field pointing toward the observer.
For PSR J1643$-$1224 there is a potential ambiguity owing to the
possible presence of orthogonal modes; the PA differnce was computed
from the first component. For the three pulsars where measurements
were available at both 20~cm and 50~cm (Table~\ref{tb:polpar}), the RM
was computed from the observed PAs at the two frequencies. The RM
increment corresponding to the $180\degr$ cycle of PA difference is
$\pm 20.1$ rad m$^2$. The value closest to the weighted mean of the
independent RMs determined at the two frequencies was chosen.

\begin{table}
\caption{Rotation measures for southern millisecond pulsars }\label{tb:rm}
\begin{tabular}{lrrrrr}
\hline
\multicolumn{1}{c}{PSR}&
\multicolumn{1}{c}{$l$}&
\multicolumn{1}{c}{$b$}&
\multicolumn{1}{c}{Dist.}&
\multicolumn{1}{c}{RM}&
\multicolumn{1}{c}{$\langle B_{||}\rangle$}
\\ 
 &
\multicolumn{1}{c}{(deg.)}&
\multicolumn{1}{c}{(deg.)}&
\multicolumn{1}{c}{(kpc)}&
\multicolumn{1}{c}{(rad m$^{-2}$)}&
\multicolumn{1}{c}{($\mu$G)}
\\
\hline
J0613$-$0200 & 210.41 &  $-9.30$ & 1.71 & $+19\pm14$  & $+0.6\pm0.4$\\
J0711$-$6830 & 279.53 & $-23.28$ & 0.86 & $+67\pm23$  & $+4.5\pm1.5$\\
J1045$-$4509 & 280.85 & $+12.25$ & 1.96 & $+82\pm18$  & $+1.7\pm0.4$\\
J1603$-$7202 & 316.63 & $-14.50$ & 1.17 & $+20.1\pm0.5$  & $+0.65\pm0.02$\\
J1623$-$2631 & 350.98 & $+15.96$ & 1.80 & $-8\pm20$   & $-0.2\pm0.4$\\
J1643$-$1224 &   5.67 & $+21.22$ & 2.41 & $-263\pm15$ & $-5.2\pm0.3$\\
J2124$-$3358 &  10.93 & $-45.44$ & 0.27 & $+1.2\pm0.1$& $+0.32\pm0.03$\\
J2129$-$5721 & 338.01 & $-43.57$ & 1.36 & $+37.3\pm0.2$   & $+1.45\pm0.01$\\
J2145$-$0750 &  47.78 & $-42.08$ & 0.57 & $+12\pm8$   & $+1.6\pm1.1$\\\hline
\end{tabular}
\end{table}

Two of these pulsars, J0711$-$6830 and J1643$-$1224, have surprisingly
large values of $|\langle B_{||}\rangle|$, especially PSR J1643$-$1224 given
its relatively large distance. In both cases, the $|\langle B_{||}\rangle|$
values are larger than for other pulsars located close to them on the
sky \citep{hmq99}. These results will be further discussed in a forthcoming
paper in the context of a large sample of pulsar RMs.

\section{Discussion and Conclusions}
In this paper we have presented pulse profiles and polarization
parameters for nine southern MSPs, five of which had no previously
published polarization data. New RM values are given for all nine
pulsars. Notable results include the very high fractional circular
polarization ($\sim 65$\% at 1327 MHz) in the trailing component of
PSR J1603$-$7202, and the extremely wide and complex pulse profile and
polarization properties of PSR J2124$-$3358. \citet{kxl+98} have argued
that, on average, MSPs have profiles which are of similar complexity
(quantified by the number of recognizable components) to those of
normal pulsars. However, there is a tail to the distribution in MSPs
exemplified by PSR J0473$-$4715 \citep{nms+97}, which has at least 10
components, and PSR J2124$-$3358 (Figure~\ref{fg:2124}; at least 12
components) which is not present in normal pulsars; no normal
pulsars have profiles as complex as these, even in data with high
signal-to-noise ratio.

While many MSP profiles have the appearance of `stretched' normal
pulsar profiles, there are significant differences. One of the most
important is the lack of frequency evolution in the spacing of pulse
components \citep{kll+99}. In most normal pulsars, the pulse
components are more widely separated at lower radio frequencies, an
observation often interpreted in terms of `radius-to-frequency
mapping' \citep[e.g.][]{cor78,tho91a}. The results presented here
confirm this lack of frequency evolution in MSP profiles. For example,
Figure~\ref{fg:2124_alpha} shows that for PSR J2124$-$3358, although
components are mostly overlapping, there is a clear correspondence of
component locations at the two frequencies right across the
profile. It is important to note however that this property is shared
by normal pulsars with so-called `interpulse' emission, including such
well-known examples as the Crab pulsar \citep{mh99} and PSR B0950+08
\citep{hc81}, suggesting a close link between the emission processes
in MSPs and `normal' interpulse pulsars.

Most MSP profiles do not fit easily into the `core-cone'
classification scheme \citep{ran83,lm88} or its extensions to multiple
cones \citep[e.g.][]{md99}. Even when the profile has a shape similar
to a classical `triple', for example PSR J2145$-$0750
(Figure~\ref{fg:2145}), the spectral properties of the central and
outer components do not follow the patterns established for normal
pulsars. For PSR J2124$-$3358 there is no identifiable `core'
component and there is no pattern suggesting multiple nested conal
emission. While there do appear to be discrete quasi-gaussian
components within the observed profile, a very large number of them
would be required to model it and their amplitude and phase
(longitude) shows no clear pattern. Rather, the profile seems more
consistent with a random distribution of emission patches as advocated
by \citet{lm88}. The spectral index variations shown in
Figure~\ref{fg:2124_alpha} are consistent with different pulse
components originating in spatially distinct emission
regions. 

\begin{figure*}
\psfig{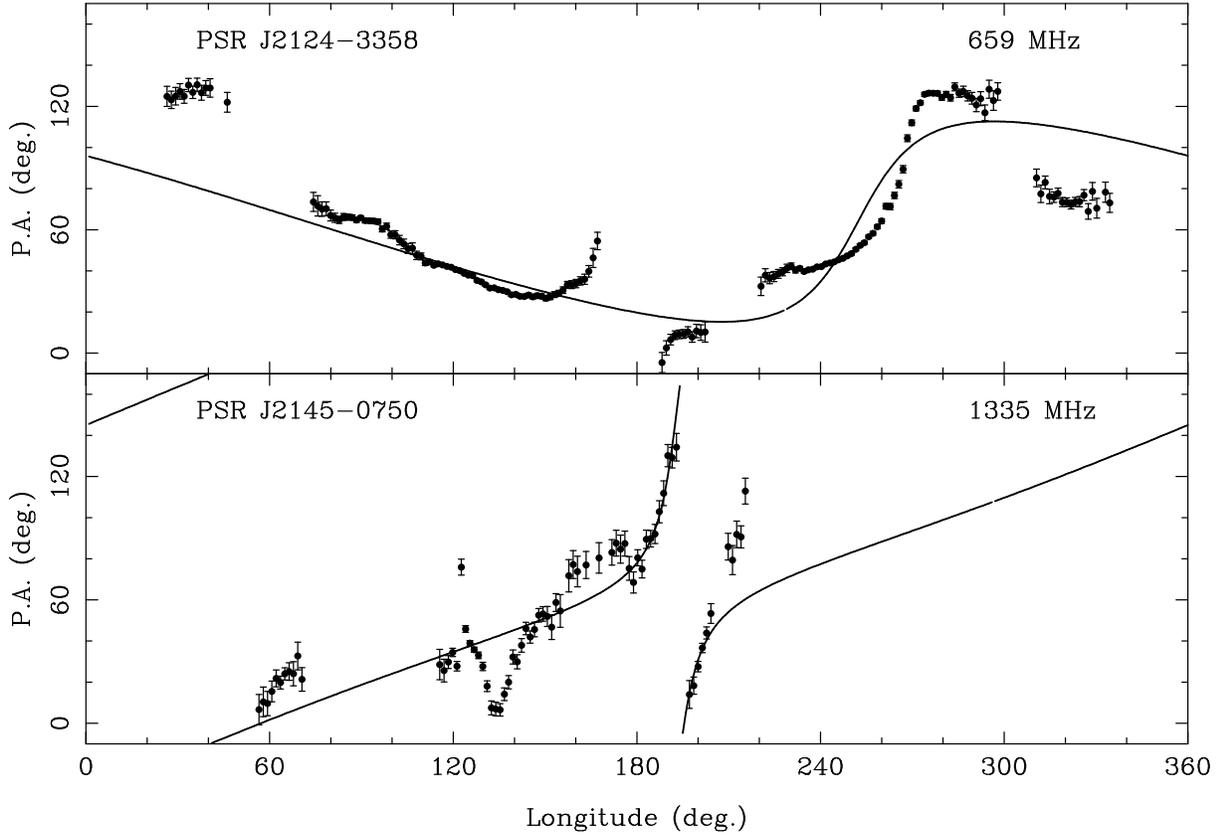}
\caption{Rotating-vector model fits to observed position-angle
variations for two pulsars.}
\label{fg:pafits}
\end{figure*}

The wide observed profiles of MSPs give a good opportunity to fit the
RVM and hence to determine both the magnetic inclination $\alpha$ and
the line-of-sight inclination $\zeta$. Most normal pulsars have
narrow profiles and only the impact parameter $\beta = \zeta - \alpha$
can be measured \citep[cf.][]{lm88}. In some MSPs, the observed PA
variations follow the RVM quite well, especially at higher frequencies
\citep[e.g.][]{stc99,lzb+00}. However, in others, there are clear and
sometimes large deviations \citep[e.g.][]{aptw96,nms+97}. Two pulsars
in the present sample, PSRs J2124$-$3358 and 2145$-$0750, have
sufficiently wide profiles to justify an attempt to fit the RVM. The
results are shown in Figure~\ref{fg:pafits}. 

While the overall PA variation for PSR J2124$-$3358 is approximately
described by the RVM fit, it is clear that there are large and
systematic deviations from the model fit. The parameters of this fit
are $\alpha = 48\degr\pm 3\degr$ and $\zeta = 67\degr\pm 5\degr$,
suggesting that emission is seen from both poles with impact
parameters $\beta = 19\degr$ and $65\degr$ respectively. However,
there is a large covariance between $\alpha$ and $\zeta$ with only a
small increase in $\chi^2$ for $\alpha$ values between $20\degr$ and
$60\degr$. At these limits, the corresponding $\zeta$ values are
$27\degr$ and $80\degr$ respectively. For small $\alpha$, a one-pole
interpretation is favoured. Given the poor quality of the fit, it is
not at all clear what the true inclination angles for the two axes
are. Certainly, the fact that emission is seen over most if not all of
the pulse period suggests a one-pole, almost aligned model.

For PSR J2145$-$0750, if we assume there is an orthogonal flip near
longitude $120\degr$ but not at $195\degr$, we get the fit shown in
Figure~\ref{fg:pafits}. This has $\alpha = 145\degr\pm 12\degr$ and
$\zeta = 148\degr\pm 12\degr$. Again, there are systematic deviations
from the fit and a large covariance between $\alpha$ and $\zeta$, but
the impact parameter $\beta \simeq 3\degr$ is relatively well
determined. This PA fit, the sense reversal of the circular
polarization and the spectral behaviour all suggest that, despite the
profile morphology, the trailing component is in fact central to the
emission beam \citep[cf.][]{stc99}. If we assume a second orthogonal
jump at $195\degr$ the impact parameter increases to about
$15\degr$. However, this introduces a significant discontinuity in PA
at this longitude, suggesting that this is not the correct
interpretation.

If we accept that the trailing component is in fact central to the
emission beam, the interpretation of the weak leading component
becomes something of a problem. \citet{xkj+98} suggest that it is a
`precursor', analogous to that in the Crab pulsar. This is supported
by its nearly 100\% linear polarization. The emission bridge between
the leading and strongest components suggests that all components are
closely related and probably that all the emission orginates from one
pole. This implies a very large effective pulse width,
$2\Delta\phi \sim 300\degr$. For small $\beta$, $\sin(\rho/2) \simeq
\sin(\Delta\phi/2)\sin(\alpha/2)$, where $\rho$ is the true beam
radius \citep[cf.][]{lm88}, giving $\rho \sim 68\degr$. This compares
with a predicted value of $49\degr$ from the relation $\rho =
6\fdg2\;P^{-1/2}$ \citep{big90b}.

For other pulsars in the sample, and indeed for most MSPs
\citep{xkj+98,stc99}, there is very little swing in PA across the the
observed pulse components.  Taken at face value, this implies large
impact parameters and/or very small magnetic inclinations. Yet many of
these pulsars have relatively narrow pulses, e.g. PSRs J0621+1002,
J1518+4904 and J1713+0747 \citep{stc99}. This implies a beam radius in
the longitude direction much less than the apparent impact parameter,
that is, a beam effectively elongated in the latitude direction
\citep[cf.][]{nv83}. If these pulse components are part of a wide beam
from magnetic field lines associated with one pole, as suggested for
young `interpulse' pulsars by \citet{man96}, intermediate magnetic
inclinations and large impact parameters are possible.  In this case,
the implied emission altitudes relative to the light cylinder are much
higher than those implied by observed pulse widths for normal
pulsars. At these altitudes, there will be significant deviations of
the magnetic field from a pure dipole form due to displacement and
physical currents, providing a possible reason for the observed large
PA deviations from RVM fits seen in millisecond pulsars. 

It is also possible that these irregular PA variations result from
very different magnetic field structures in MSPs. For example,
\citet{rud91b} suggests that, during the spindown and subsequent
recycling process, crustal plate movements result in highly distorted
field structures. Accretion-induced field decay \citep[e.g.][]{rom90}
would also result in highly non-dipolar fields. Such field structures
could also disguise or destroy the relatively consistent patterns of
core and conal emission seen in most longer-period pulsars.

In general, our polarization results agree well with those of
\citet{stc99} when account is taken of the reversed PA convention in
that paper. However, there are many discrepancies with results in the
\citet{xkj+98} paper. It appears that their sign of Stokes $V$ is
systematically reversed. It is also probable that the higher and
approximately constant fractional linear polarization with nearly
constant PA seen in a number of their figures results from errors in
calibration rather than time variations in the polarization properties
\citep[cf.][]{stc99}.

\section*{Acknowledgments}

We thank the referee, Joanna Rankin, for helpful comments. JLH thanks
the exchange program between CAS and CSIRO for support of the visits
at the Australia Telescope National Facility in 1999 and 2000.  His
research is supported by the National Natural Science Foundation of
China (10025313) and the National Key Basic Research Science
Foundation of China (G19990756).  The Australia Telescope is funded by
the Commonwealth Government for operation as a National Facility
managed by CSIRO.


\begin{thebibliography}{43}
\expandafter\ifx\csname natexlab\endcsname\relax\def\natexlab#1{#1}\fi

\bibitem[{Arzoumanian {et~al.}(1996)Arzoumanian, Phillips, Taylor, \&
  Wolszczan}]{aptw96}
Arzoumanian, Z., Phillips, J.~A., Taylor, J.~H., \& Wolszczan, A. 1996, ApJ,
  470, 1111

\bibitem[{Backer(1976)}]{bac76}
Backer, D.~C. 1976, ApJ, 209, 895

\bibitem[{Bailes {et~al.}(1997)Bailes, Johnston, Bell, Lorimer, Stappers,
  Manchester, Lyne, D'Amico, \& Gaensler}]{bjb+97}
Bailes, M., Johnston, S., Bell, J.~F., Lorimer, D.~R., Stappers, B.~W.,
  Manchester, R.~N., Lyne, A.~G., D'Amico, N., \& Gaensler, B.~M. 1997, ApJ,
  481, 386

\bibitem[{Bell {et~al.}(1997)Bell, Bailes, Manchester, Lyne, Camilo, \&
  Sandhu}]{bbm+97}
Bell, J.~F., Bailes, M., Manchester, R.~N., Lyne, A.~G., Camilo, F., \& Sandhu,
  J.~S. 1997, MNRAS, 286, 463

\bibitem[{Biggs(1990)}]{big90b}
Biggs, J.~D. 1990, MNRAS, 245, 514

\bibitem[{Cordes(1978)}]{cor78}
Cordes, J.~M. 1978, ApJ, 222, 1006

\bibitem[{{Cordes} \& Lazio(2002)}]{cl02}
{Cordes}, J.~M. \& Lazio, T.~J.~W. 2002,
  http://xxx.lanl.gov/abs/astro-ph/0207156

\bibitem[{Gould \& Lyne(1998)}]{gl98}
Gould, D.~M. \& Lyne, A.~G. 1998, MNRAS, 301, 235

\bibitem[{{Han} \& {Manchester}(2001)}]{hm01}
{Han}, J.~L. \& {Manchester}, R.~N. 2001, MNRAS, 320, L35

\bibitem[{Han {et~al.}(2002)Han, Manchester, Lyne, \& Qiao}]{hmlq02}
Han, J.~L., Manchester, R.~N., Lyne, A.~G., \& Qiao, G.~J. 2002, ApJ, 570, L17

\bibitem[{{Han} {et~al.}(1999){Han}, {Manchester}, \& {Qiao}}]{hmq99}
{Han}, J.~L., {Manchester}, R.~N., \& {Qiao}, G.~J. 1999, MNRAS, 306, 371

\bibitem[{Han {et~al.}(1998)Han, Manchester, Xu, \& Qiao}]{hmxq98}
Han, J.~L., Manchester, R.~N., Xu, R.~X., \& Qiao, G.~J. 1998, MNRAS, 300, 373

\bibitem[{Hankins \& Cordes(1981)}]{hc81}
Hankins, T.~H. \& Cordes, J.~M. 1981, ApJ, 249, 241

\bibitem[{{Helfand} {et~al.}(2001){Helfand}, {Gotthelf}, \& {Halpern}}]{hgh01}
{Helfand}, D.~J., {Gotthelf}, E.~V., \& {Halpern}, J.~P. 2001, ApJ, 556, 380

\bibitem[{Komesaroff(1970)}]{kom70}
Komesaroff, M.~M. 1970, Nature, 225, 612

\bibitem[{Kramer {et~al.}(1999{\natexlab{a}})Kramer, Lange, Lorimer, Backer,
  Xilouris, Jessner, \& Wielebinski}]{kll+99}
Kramer, M., Lange, C., Lorimer, D.~R., Backer, D.~C., Xilouris, K.~M., Jessner,
  A., \& Wielebinski, R. 1999{\natexlab{a}}, ApJ, 526, 957

\bibitem[{Kramer {et~al.}(1994)Kramer, Wielebinski, Jessner, Gil, \&
  Seiradakis}]{kwj+94}
Kramer, M., Wielebinski, R., Jessner, A., Gil, J.~A., \& Seiradakis, J.~H.
  1994, A\&AS, 107, 515

\bibitem[{Kramer {et~al.}(1999{\natexlab{b}})Kramer, Xilouris, Camilo, Nice,
  Lange, Backer, \& Doroshenko}]{kxc+99}
Kramer, M., Xilouris, K.~M., Camilo, F., Nice, D., Lange, C., Backer, D.~C., \&
  Doroshenko, O. 1999{\natexlab{b}}, ApJ, 520, 324

\bibitem[{Kramer {et~al.}(1998)Kramer, Xilouris, Lorimer, Doroshenko, Jessner,
  Wielebinski, Wolszczan, \& Camilo}]{kxl+98}
Kramer, M., Xilouris, K.~M., Lorimer, D.~R., Doroshenko, O., Jessner, A.,
  Wielebinski, R., Wolszczan, A., \& Camilo, F. 1998, ApJ, 501, 270

\bibitem[{{Lommen} {et~al.}(2000){Lommen}, {Zepka}, {Backer}, {McLaughlin},
  {Cordes}, {Arzoumanian}, \& {Xilouris}}]{lzb+00}
{Lommen}, A.~N., {Zepka}, A., {Backer}, D.~C., {McLaughlin}, M., {Cordes},
  J.~M., {Arzoumanian}, Z., \& {Xilouris}, K. 2000, ApJ, 545, 1007

\bibitem[{Lorimer {et~al.}(1996)Lorimer, Lyne, Bailes, Manchester, D'Amico,
  Stappers, Johnston, \& Camilo}]{llb+96}
Lorimer, D.~R., Lyne, A.~G., Bailes, M., Manchester, R.~N., D'Amico, N.,
  Stappers, B.~W., Johnston, S., \& Camilo, F. 1996, MNRAS, 283, 1383

\bibitem[{Lyne \& Manchester(1988)}]{lm88}
Lyne, A.~G. \& Manchester, R.~N. 1988, MNRAS, 234, 477

\bibitem[{Manchester(1996)}]{man96}
Manchester, R.~N. 1996, in Pulsars: Problems and Progress, {IAU} Colloquium
  160, ed. S.~Johnston, M.~A. Walker, \& M.~Bailes (San Francisco: Astronomical
  Society of the Pacific), 193--196

\bibitem[{Mitra \& Deshpande(1999)}]{md99}
Mitra, D. \& Deshpande, A.~A. 1999, A\&A, 346, 906

\bibitem[{Moffett \& Hankins(1999)}]{mh99}
Moffett, D.~A. \& Hankins, T.~H. 1999, ApJ, 522, 1046

\bibitem[{Narayan \& Vivekanand(1983)}]{nv83}
Narayan, R. \& Vivekanand, M. 1983, A\&A, 122, 45

\bibitem[{Navarro(1994)}]{nav94}
Navarro, J. 1994, PhD thesis, California Institute of Technology

\bibitem[{Navarro {et~al.}(1997)Navarro, Manchester, Sandhu, Kulkarni, \&
  Bailes}]{nms+97}
Navarro, J., Manchester, R.~N., Sandhu, J.~S., Kulkarni, S.~R., \& Bailes, M.
  1997, ApJ, 486, 1019

\bibitem[{Radhakrishnan \& Cooke(1969)}]{rc69a}
Radhakrishnan, V. \& Cooke, D.~J. 1969, Astrophys. Lett., 3, 225

\bibitem[{Radhakrishnan \& Deshpande(2001)}]{rd01}
Radhakrishnan, V. \& Deshpande, A.~A. 2001, A\&A, 379, 551

\bibitem[{Rankin(1983)}]{ran83}
Rankin, J.~M. 1983, ApJ, 274, 333

\bibitem[{Rankin(1990)}]{ran90}
---. 1990, ApJ, 352, 247

\bibitem[{Rankin(1993)}]{ran93}
---. 1993, ApJ, 405, 285

\bibitem[{Rankin \& Rathnasree(1997)}]{rr97}
Rankin, J.~M. \& Rathnasree, N. 1997, J. Astrophys. Astr., 18, 91

\bibitem[{Romani(1990)}]{rom90}
Romani, R.~W. 1990, Nature, 347, 741

\bibitem[{Ruderman(1991)}]{rud91b}
Ruderman, M. 1991, ApJ, 382, 576

\bibitem[{Stairs {et~al.}(1999)Stairs, Thorsett, \& Camilo}]{stc99}
Stairs, I.~H., Thorsett, S.~E., \& Camilo, F. 1999, ApJS, 123, 627

\bibitem[{Staveley-Smith {et~al.}(1996)Staveley-Smith, Wilson, Bird, Disney,
  Ekers, Freeman, Haynes, Sinclair, Vaile, Webster, \& Wright}]{swb+96}
Staveley-Smith, L., Wilson, W.~E., Bird, T.~S., Disney, M.~J., Ekers, R.~D.,
  Freeman, K.~C., Haynes, R.~F., Sinclair, M.~W., Vaile, R.~A., Webster, R.~L.,
  \& Wright, A.~E. 1996, Proc. Astr. Soc. Aust., 13, 243

\bibitem[{Stinebring {et~al.}(1984)Stinebring, Cordes, Rankin, Weisberg, \&
  Boriakoff}]{scr+84}
Stinebring, D.~R., Cordes, J.~M., Rankin, J.~M., Weisberg, J.~M., \& Boriakoff,
  V. 1984, ApJS, 55, 247

\bibitem[{Thorsett(1991)}]{tho91a}
Thorsett, S.~E. 1991, ApJ, 377, 263

\bibitem[{Thorsett \& Stinebring(1990)}]{ts90}
Thorsett, S.~E. \& Stinebring, D.~R. 1990, ApJ, 361, 644

\bibitem[{Toscano {et~al.}(1998)Toscano, Bailes, Manchester, \&
  Sandhu}]{tbms98}
Toscano, M., Bailes, M., Manchester, R., \& Sandhu, J. 1998, ApJ, 506, 863

\bibitem[{Xilouris {et~al.}(1998)Xilouris, Kramer, Jessner, von Hoensbroech,
  Lorimer, Wielebinski, Wolszczan, \& Camilo}]{xkj+98}
Xilouris, K.~M., Kramer, M., Jessner, A., von Hoensbroech, A., Lorimer, D.,
  Wielebinski, R., Wolszczan, A., \& Camilo, F. 1998, ApJ, 501, 286

\end{thebibliography}

\end{document}